\documentclass[%
 reprint,
 amsmath,amssymb,
 aip,
]{revtex4-2}
\usepackage{graphicx}
\usepackage{dcolumn}
\usepackage{bm}
\usepackage{xcolor}

\usepackage[load=physical,load=synchem]{siunitx}
\DeclareSIUnit\electron{e^-}

\newcommand{\FI}{\mathcal{I}}
\newcommand{\QFI}{\mathcal{J}}
\newcommand{\mean}[1]{\left\langle#1\right\rangle}

\newcommand{\ket}[1]{\left|#1\right\rangle}
\newcommand{\bra}[1]{\left\langle#1\right|}
\newcommand{\braket}[2]{\left\langle\left. #1\right| #2\right\rangle}

\begin{document}
\preprint{APS/123-QED}

\title{Transmission Electron Microscopy at the Quantum Limit}
\author{Stewart A. Koppell}
\affiliation{%
 Physics Department, Stanford University, Stanford, California 94305, USA
}%
\affiliation{%
Department of Electrical Engineering and Computer Science, Massachusetts Institute of Technology
}%
\author{Yonatan Israel}
\author{Adam J. Bowman}
\author{Brannon B. Klopfer}

\affiliation{%
 Physics Department, Stanford University, Stanford, California 94305, USA
}%

\author{Mark A. Kasevich}%
 \email[]{Authors to whom correspondence should be addressed: kasevich@stanford.edu}
\affiliation{%
 Physics Department, Stanford University, Stanford, California 94305, USA
}%

\begin{abstract}
A number of visions for a new generation of dose-efficient electron microscopes have been advanced. These proposals, while inspired by quantum principles, make little contact with the broader field of quantum metrology. We discuss a framework calculating the amount of information carried by each electron. This makes it possible to evaluate the potential effectiveness of any particular microscope architecture relative to the quantum limit for information per dose.  In the case of phase imaging, we argue this limit is at least an order of magnitude beyond what is possible with aberration-free Zernike phase contrast.
\end{abstract}
\maketitle

\section{Introduction}
Transmission electron microscopes (TEMs) are versatile tools, essential to many fields of science, from material science to structural biology. While modern TEMs are capable of atomic resolution, a fundamental limitation remains: many samples are degraded by high energy electrons. For example, cryo-electron microscopy (cryo-EM) makes it possible to image biological molecules in their (ostensibly) native configurations\cite{dubochet1988cryo,frank2006,cryoEMStrucBioreview2015}, but these molecules are destroyed long before a high resolution image can be formed. It is possible to form a near-atomic-resolution reconstruction of the molecule using a technique called single particle analysis (SPA), whereby a large number of low resolution images are collected and then combined using a sophisticated algorithm. However this laborious procedure can only be performed on large ensembles of identical particles heavier than about $40$ kDa\cite{SPA}. Imaging of single protein instances can provide valuable insights into structural biology beyond SPA \cite{rickgauer2017, rickgauer2020}, but current techniques are constrained by the dose limit of TEM.
A typical dose for cryo-EM is tens of electrons per square angstrom, and is often divided into multiple exposures. With such limited measurement resources, it is important to extract as much information as possible from each electron. 

Quantum metrology is dedicated to pursuing the ultimate sensitivity of measurements allowed by quantum mechanics. This is made possible by developing formalism to calculate such limits and developing experimental means to achieve them in practice. In this perspective we comment on research that lies in the intersection of the two fields, whereby tools of quantum metrology are used to achieve the ultimate limits allowed by quantum mechanics in electron microscopy. 

\section{Visions of a Quantum Electron Microscope}

In the context of quantum metrology, a ‘quantum’ enhanced measurement is generally one which uses a non-classical probe state to achieve lower error per measurement resource. Arguably the most important result of quantum metrology is that by using a probe of $n$ particles prepared in an entangled state, it is possible to reduce measurement variance by a factor scaling like $n^2$ (the Heisenberg Limit) rather than like $n$ (the Standard Quantum Limit)\cite{QuantumMZI_2002,QuantMeas_2004}. However, Heisenberg-limited measurements are only possible for lossless system\cite{QL}. 
In damage-limited TEM, where loss is intrinsic, what is the benefit of a quantum treatment? We will argue that there is a quantum limit for information per damage which an ideal, traditional TEM architecture cannot access. We will call any microscope which approaches this limit a Quantum Electron Microscope (QEM).

\begin{figure*}
    \centering
    \includegraphics[width=\textwidth]{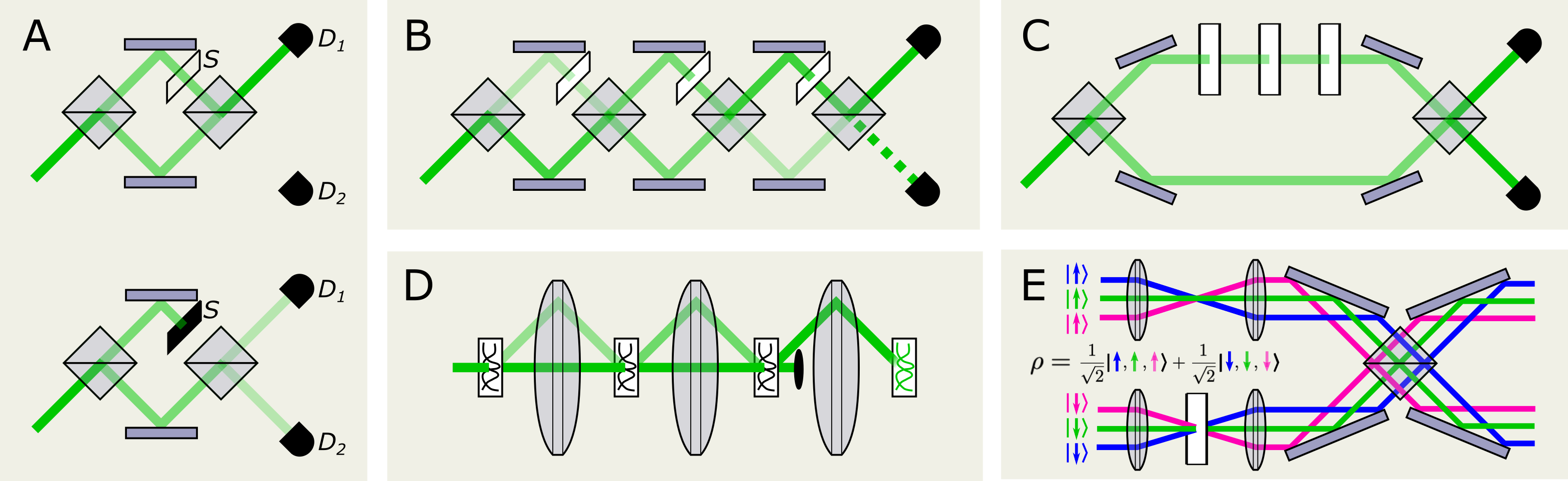}
    \caption{(A) Elitzur-Vaidman bomb test with transparent (top) or opaque (bottom) sample $S$. (B) Elitzur-Vaidman bomb test with weak, repeated interrogations. This configuration is known as interaction-free measurement (IFM). (C) Multi-pass (MP) interferometry. (D) MP microscopy. For simplicity, the schematics in (B), (C), and (D) show multiple copies of the sample. In practice these measurements can be performed with a single copy by forming the optical axis into a loop. (E) Interferometer scheme compatible with entanglement-enhanced fermionic probes. The Pauli-exclusion principle is circumvented by using inter-mode correlations.}
    \label{fig:IFM-MP}
\end{figure*}

In optical microscopy, entanglement-enhanced measurements have been shown to increase imaging sensitivity and reduce damage \cite{multiphaseEstPRL13,NOONmicroscopyPRL14,QuantumenhancedNM}. In principle, fermionic probes can also be enhanced using entanglement (and indeed, achieve the Heisenberg limit) by partitioning states with $n$ electrons into $k$ modes (where $k>n$)\cite{FermionNOON_PRA14}. There has been no experimental demonstration of entanglement between free electrons, but free electrons have been entangled with other systems\cite{Zhao2021,Hayun2021}. One proposal for entanglement-enhanced TEM suggests using a flux qubit to coherently sum the phase shifts measured by a train of single electron pulses\cite{Okamoto2013}. 

There have also been proposals which invoke quantum principles other than entanglement, such as interaction-free measurement\cite{ElitzurVaidman1993} (IFM). The simplest kind of IFM can be performed using a balanced interferometer as depicted in Fig. \ref{fig:IFM-MP}(A). The goal is to determine whether an opaque object ($S$) has been placed in one of the interferometer arms. When $S$ is absent, the interferometer is configured to have complete constructive interference at one output port ($D_1$) and destructive interference at the other ($D_2$). When $S$ is present, there is a 50\% chance that it will absorb each photon (a failed IFM) and a 25\% chance that the photon arrives at $D_2$. A single photon detection at $D_2$ unequivocally establishes the presence of $S$ while also precluding interaction\cite{Vaidman2001} with $S$.  The success probability can be increased arbitrarily close to unity using an asymmetric interferometer to repeatedly (weakly) interrogate the sample\cite{kwiat1995} as shown in Fig \ref{fig:IFM-MP}(B). IFM can be thought of in terms of the quantum Zeno effect\cite{misra1977zeno,quantumZeno1990,kwiat1995}, where the obstructing object acts as a measuring device to freeze the coherent evolution of the probe. 

The first proposal for IFM in electron microscopy called for a scanning probe of \SI{100}{\kilo\electronvolt} electrons confined in a hybrid Paul-Penning trap\cite{Putnam2009}. The scheme relied on sample staining (with heavy atoms) to produce the high contrast necessary for efficient IFM, and predicted a reduction in sample damage by two orders of magnitude. In light of subsequent analysis, this seems overly optimistic\cite{Thomas2014}. Furthermore, the high-contrast requirement excludes the most important applications for a QEM (e.g., imaging weak phase objects like proteins in vitreous ice). A later proposal was based on a more traditional TEM architecture with the addition of a coherent beam-splitter and `resonator' (a re-imaging cavity with fast in/out-coupling optics) to allow multiple passes through the sample\cite{og_qemdesign}. It was also suggested that IFM could be used to measure a phase shift, but the proposal did not include a calculation of the efficiency of this technique relative to standard TEM. Recently, IFM was demonstrated for electrons with an efficiency of 14\% using diffraction grating based beam-splitters [Fig. \ref{fig:IFM-MP}(A)] in a balanced interferometer \cite{E-IFM2021}.

A well-known method for enhancing sensitivity in measurements of a static phase shifts involves simply passing the probe through the sample $m$ times before detection [see Fig. \ref{fig:IFM-MP}(C)]. Such sequential or multi-pass (MP) measurements have been shown to be quantum-optimal\cite{MPoptimality2,MPoptimality1} and proposed as a substitution for entanglement in interferometric measurements\cite{MPoptimality3,MPinterferometer}. The idea can also be applied to full-field imaging, where it is known as MP microscopy\cite{MP_NatComm2016,MPsim_SciRep2017}.
In this technique [see Fig. \ref{fig:IFM-MP}(D)], a probe is re-imaged $m$ times onto the specimen, thereby enhancing the phase shift accrued by a factor of $m$. This results in an increase of the signal-to-noise ratio (SNR) by a factor of $m$, or by a factor $\sqrt{m}$ at constant damage. Unlike IFM, a MP microscope does not require a beam-splitter. Nevertheless, for weakly scattering samples, there is a strong physical analogy between full-field MP and interaction-free measurements. The specimen in a MP microscope heuristically acts like a beam-splitter, scattering and combining the beam with each pass. The result is an increase in the probability of detecting the presence of a scatterer (through an elastic interaction) relative to the probability of causing damage (through an inelastic interaction).

There have also been proposed solutions to a more specialized measurement scenario where the goal is to identify the sample from a finite list of structural hypotheses. The corresponding problem in quantum metrology is called quantum state discrimination (QSD). In the simplest case, where the position and orientation of the sample are known and losses are negligible, it is possible to distinguish between two phase objects in a single shot using an entangled probe\cite{NOONdiscrimination} or a MP measurement with a single probe particle\cite{MPdiscrimination}. The key to applying QSD to TEM is inventing optics for projecting into an optimal measurement basis. When the position and orientation are known, a single phase mask may be sufficient\cite{Okamoto2006}. When the orientation is unknown, efficient discrimination is still possible by using a sequence of diffractive elements to measure in an orbital angular momentum basis \cite{qsort}. An efficient strategy for QSD when both the position and orientation are unknown has yet to be proposed.

A number of projects are underway to turn some of the conceptual innovations described in this section into proof-of-principle devices. How can we know whether any of these schemes have the potential to saturate the quantum limit for dose-limited information? In the next section, we map the problem of phase imaging in TEM onto phase estimation in quantum metrology.

\section{Electron Microscopy as a Quantum Metrology Protocol}
\label{sec:QFI}
The paradigmatic approach to quantum metrology begins by describing the unknown degrees of freedom in the sample using a parameter vector $\Theta=[\theta_0,\ldots,\theta_n]$. If the purpose of the measurement is to image a completely general phase object, the parameter values could represent the projected atomic potential $V(\mathbf{r})=\int dz V(z,\mathbf{r})$ at each transverse coordinate $\mathbf{r}$ or the amplitude of the Fourier coefficients of $V$ for each spatial frequency $\mathbf{k}$. It is often possible to reduce the dimension of $\Theta$ using prior information about the sample. For single particle analysis, $\Theta$ may describe the species and coordinates of each atom in a model particle and the positions and orientations of each (hypothetically identical) particle in the field of view. When measuring summary material properties like a lattice constant, average grain size, or vacancy populations, $\Theta$ may be scalar.

A metrological protocol for estimating $\Theta$ consists of four steps. 
\begin{enumerate}
    \item Create a probe in (mixed) state $\rho$.
    \item Interact the probe with the sample represented by transmission function $S_\Theta$ to produce output state $\rho_\Theta=S_\Theta(\rho)$.
    \item Project the output state onto a set of $d$ detectors using transfer function $T$ to produce intensities $I_j=\bra{j}T(\rho_\Theta)\ket{j}$.
    \item Analyze $I$ using a set of estimating functions (estimators) $\hat{{\Theta}}$.
\end{enumerate}
The standard way to formally optimize this protocol is by minimizing a cost function 
\begin{equation}
C=\mathbf{w}\cdot\text{Var}(\hat{{\Theta}})=\sum_i w_i \text{Var}(\hat{\theta}_i)\ ,
\end{equation}
where the weights $w_i$ define the relative value of information about each parameter. We may use a non-uniform weighting, for example, to prioritize the high spatial-frequency information which tends to degrade more quickly as a result of radiation damage.

The information $\FI_i$ contained in $I$ about $\theta_i$ is related to the minimum measurement variance achieved using an optimal (and unbiased) estimator. Similarly, the information $\QFI_i$ available in $\rho_\Theta$ about $\theta_i$ is related to the maximum $\FI_i$ achieved by using an optimal transfer function. The quantities $\FI_i$ and $\QFI_i$ are called the Fisher Information (FI) and Quantum Fisher Information (QFI)\cite{Holevo1982,Helstrom1967,Braunstein1994} respectively. They are related to the measurement variance by the Cram\'er-Rao Bound (CRB) and the Quantum CRB (QCRB)
\begin{equation}
    \text{Var}(\hat{\theta}_i)\underset{\text{CRB}}{\geq} \frac{1}{N\FI_i(\Theta,\rho,T)}\underset{\text{QCRB}}{\geq} \frac{1}{N\QFI_i(\Theta,\rho)}\ ,
    \label{eq:CRB}
\end{equation}
where $N$ is the number of independent measurements. It is worth noting that the FI and QFI have units of inverse variance rather than bits. However, like the more traditional notion of information (i.e., entropy) and unlike figures of merit such as contrast and resolution, the FI and QFI from independent measurements are additive. The utility of this formalism comes from explicit formulae for the FI and QFI, which establish optimization benchmarks for $\hat{\Theta}$ and $T$, respectively.

Given a model for $I$ as a function of $\Theta$, the FI is
\begin{equation}
    \FI_i=\sum_{j=1}^d \frac{1}{I_j}\left(\frac{\partial}{\partial\theta_i}I_j\right)^2
\end{equation}
and in the special case that the probe is a pure state $\rho_\Theta=\psi$, the QFI is
\begin{equation}
    \QFI_i=4\braket{\frac{\partial}{\partial\theta_i}\psi}{\frac{\partial}{\partial\theta_i}\psi}-4\left|\braket{\frac{\partial}{\partial\theta_i}\psi}{\psi}\right|^2 \ .
    \label{eq:QFI}
\end{equation}
As an example, consider a sample with transmission function $S_\Theta(\rho)=\exp\left(i\phi(\mathbf{r})\right)\ket{\mathbf{r}}\bra{\mathbf{r}}\rho$ where $\phi(\mathbf{r})=aV(\mathbf{r})$, $a$ is an interaction constant depending on the beam energy (derivable from the WKB approximation), and $V(\mathbf{r})=\sum_{\mathbf{k}}\theta_{\mathbf{k}} \cos(\mathbf{r}\cdot\mathbf{k})$. Suppose we use a pure, plane wave probe and the projection optics are described by transfer function $T(\rho)=\mathcal{F}^{-1}\left\{\exp\left(i\chi(|\mathbf{k}|)\right)\mathcal{F}\{\rho\}\right\}$ which implements a radially-symmetric phase mask in the back focal plane of the objective lens. In the weak phase object approximation (WPOA, $\phi(\mathbf{r})\ll1$)
\begin{equation}
    \FI_i=4a^2\sin^2(\chi(|\mathbf{k}_i|)-\chi(0))=4a^2C(|\mathbf{k}_i|)^2\ ,
\end{equation}
where $C(|\mathbf{k}|)$ is the phase contrast transfer function (CTF) as it is typically defined for TEM\cite{CTF}. The QFI can be calculated using equation \ref{eq:QFI} or by optimizing $\chi$ by inspection. Regardless of the weighting $\mathbf{w}$, the optimal transfer function has $\chi(|\mathbf{k}|)-\chi(0)=\pi/2$ for all $k$, corresponding to Zernike phase contrast (ZPC) and $\QFI_i=4a^2$. 

When the WPOA cannot be applied, each spatial frequency component of $V$ may generate contrast at many spatial frequencies at the detector (i.e., a strong phase grating scatters to many angles). In this case, to account for possible covariance between measurements of $\theta_i$ and $\theta_j$, the FI and QFI can be formulated as matrices $\FI_{i,j}$ and $\QFI_{i,j}$. Then CRB and QCRB provide lower bounds for the covariance matrix $\Sigma(\hat{\Theta})$. In general, the optimal transfer function depends on $\mathbf{w}$ and the particular value of $\Theta$. For example, a Zernike phase plate produces negligible contrast when the majority of the probe is scattered.

While ZPC produces an apparently ideal CTF in the WPOA, more information is attainable using more general measurements. If the probe is passed through the sample $m$ times or includes $n$ maximally entangled particles, the QFI can improve by a factor as large as $m^2$ or $n^2$ (or both can be done at once)\cite{MPoptimality1}. In principle, the information per probe particle is unbounded, however the initial uncertainty in the sample phase is also amplified, eventually causing $2\pi$ ambiguities. In practice, $m$ and $n$ should be initially chosen to avoid this, and can be increased as the variance of intermediate estimates of the phase decreases.

So far our model has not included loss and damage. As we will show, the total loss will generally affect the optimal measurement parameters (e.g., $m$ or $n$). Damage, which is associated with inelastic scattering in the sample, will limit the total number of probes $N$. Some inelastic interactions preserve partial spatial coherence in the probe (e.g., plasmon scattering)\cite{CTF,schattschneider200Inelcoherence,herring2005EELShologratphy,egerton2017EELSloc} so they retain some value for phase estimation. However, for highly localized interactions (e.g., chemical bond breaking), the sample presumably receives enough which-way information to destroy all interference at the detector. For simplicity, we will assume that inelastically scattered electrons contain no information about the sample phase and can be filtered from the imaging beam before they arrive at the detector or return to the sample (in a MP measurement). With this assumption, inelastic scattering in electron microscopy is similar to absorption in photon microscopy.

For measurements of a single phase shift with transmissivity $\eta$, MP and NOON state measurements are optimal (producing the maximum QFI) at fixed $n\times m$ when $\eta$ is high\cite{lossyQE}. However, with damage as the measurement resource, MP and NOON states are sub-optimal. It has been shown that for the measurement of a lossy phase shift using photons, the QFI per absorbed photon is bounded by\cite{Birchall2017}
\begin{equation}
    \xi\leq\xi_{\text{QL}}=\frac{4\eta}{(1-\eta)^2}\ .
    \label{eq:IpDQ}
\end{equation}
The calculation of $\xi_{\text{QL}}$ does not rely on bosonic particle statistics\cite{QL}, so it also bounds measurements performed with electrons (note the bound applying to an estimate of $V=\phi/a^2$ is scaled by $a^2$). The bound can be saturated by squeezed states\cite{Birchall2017} (which can be formed with fermionic probes using inter-mode entanglement\cite{FermionSqueezing}). Due to the great technical challenge of forming such states using free electrons, it is worth considering whether other measurements can approach this bound.

\begin{figure}\centering
\includegraphics[width=0.5\textwidth]{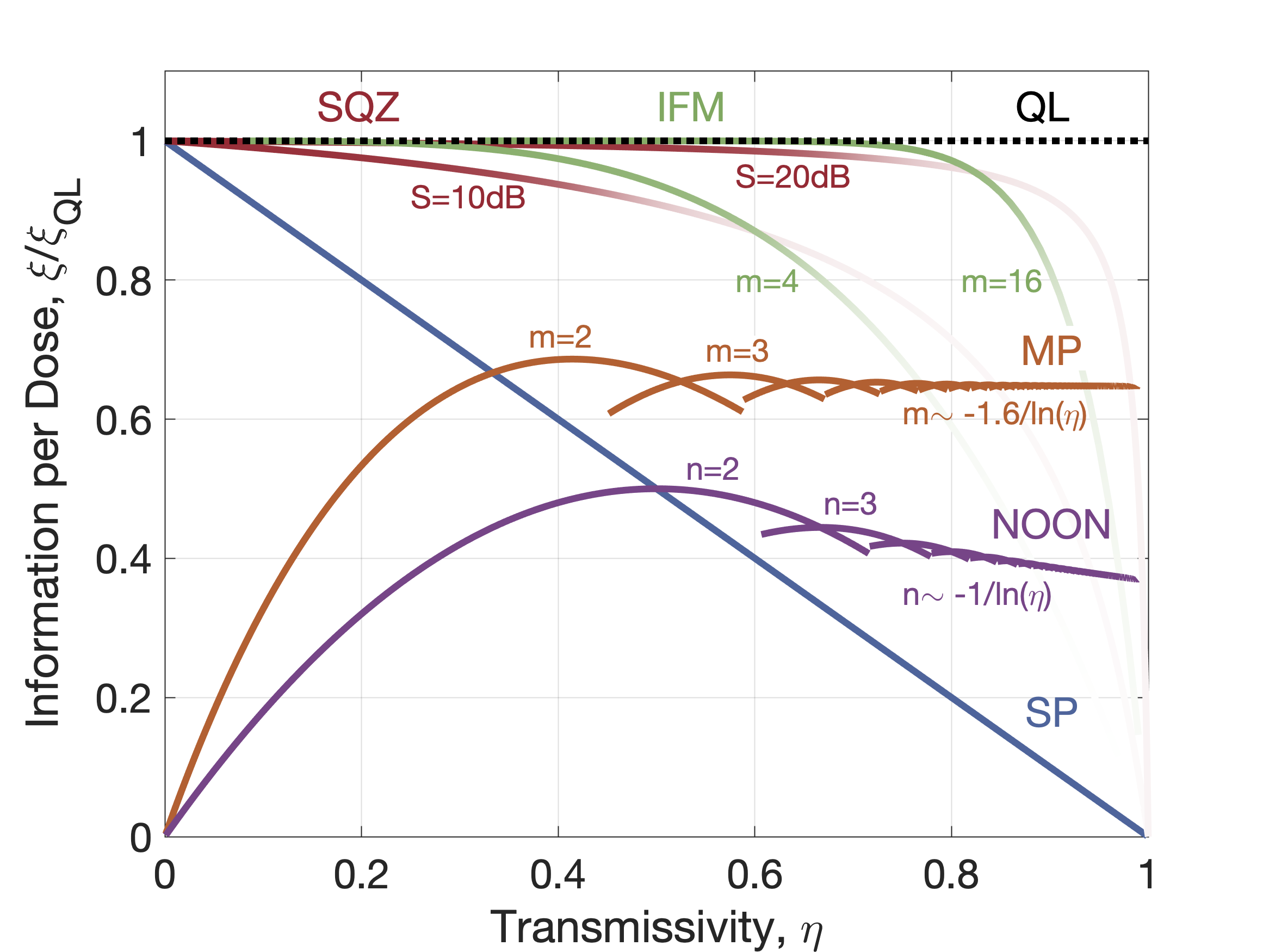}

\caption{Fisher information per dose, $\xi$, relative to the quantum limit $\xi_{\text{QL}}$ for estimation of a single lossy phase shift. At high transmissivity $\eta$, a single-pass measurement with a classical probe (SP) attains only a small fraction of the information available from more general measurements. Multi-pass (MP) and NOON state measurements offer a significant improvement when the number of passes ($m$) or probe particles ($n$) is optimized. The quantum limit can be saturated using a squeezed gaussian probe (SQZ) or using a sequence of weak interactions with a single-particle probe (IFM).}
\label{fig:QL}
\end{figure}

MP and NOON state measurements produce the same amount of information when $n=m$, however they inflict a different dose on the sample. If all loss is due to inelastic scattering in the sample, the dose is $1-\eta^m$ for a MP measurement and $n(1-\eta)$ for a NOON state measurement. When $\eta\sim 1$, $\xi_{\text{NOON}}\lesssim\xi_{\text{QL}}/e=0.38\xi_{\text{QL}}$ for optimal $n$ and $\xi_{\text{MP}}\lesssim0.65\xi_{\text{QL}}$ for optimal $m$. 
An IFM architecture can be be used to perform a still more efficient measurement. While the standard scheme uses a beam-splitter with transmissivity $\tau=\sin(\pi/2m)^2$, we have found the dose efficiency is improved by setting $\tau\ll (\pi/2m)^2$, in which case $\xi_{\text{IFM}}\gtrsim.9\xi_{\text{QL}}$. By changing $\tau$ each pass according to 
\begin{equation}
    \tau_k=\epsilon \eta^{m-k}
\end{equation}
with $\epsilon\ll1$, IFM can saturate the quantum limit. Fig. \ref{fig:QL} shows $\xi/\xi_{\text{QL}}$ for each of these measurements as a function of $\eta$.

\begin{table}
\begin{center}
\begin{tabular}{c | c | c | c | c | c }
  & Single Pass & NOON & Multi-Pass & Squeezed\\ 
 \hline
  $\QFI$ & $4\eta a^2$ & $4\eta n^2a^2$ & $4\eta m^2a^2$ & $\dfrac{4\eta\alpha^2a^2}{1-\eta+\eta 10^{-S/10}}$ \\
   \hline
  dose & $1-\eta$ & $n(1-\eta)$ & $1-\eta^m$ & $\alpha^2(1-\eta)$ \\
  \hline
   $\xi(\eta=0.9)$ & 36 & 3.9$\xi_{\text{SP}}$ & 6.4$\xi_{\text{SP}}$ & $\leq10\xi_{\text{SP}}$\\
   $\xi(\eta=0.95)$ & 76 & 7.5$\xi_{\text{SP}}$ & 13$\xi_{\text{SP}}$ & $\leq20\xi_{\text{SP}}$\\
  \hline
     $r(\eta=0.9)$ & \SI{20}{\angstrom} & \SI{10}{\angstrom} & \SI{7.9}{\angstrom} & $\geq$\SI{6.3}{\angstrom}\\
   $r(\eta=0.95)$ & \SI{20}{\angstrom} & \SI{7.7}{\angstrom} & \SI{5.5}{\angstrom} & $\geq$\SI{4.5}{\angstrom}\\
  \hline
\end{tabular}
\end{center}
  \label{tab:comparison}
  \caption{Formulae for the QFI, $\QFI$, and dose for various measurements of a single phase shift. The NOON state measurement uses $n$ particles, the multi-pass measurement uses $m$ passes, and the squeezed state has squeezing $S$ (in dB) and expects $a^2$ particles to pass through the sample. Numerical values for the QFI per dose, $\xi$, and a typical\cite{2nmRes} dose-limited resolution $r$ are provided for $\eta=0.9$ and $\eta=0.95$. }
\end{table}
In order to ground this discussion in quantities more familiar to electron microscopy, we can write $\xi$ in terms of elastic and inelastic scattering cross sections. We will focus on MP measurements, which would likely be the simplest to implement. Using the WPOA, the probability of elastic scattering is $tn_s\sigma_e^{(m)}=m^2a^2\mean{\Delta V^2}$ where $\mean{\Delta V^2}$ is the spatial variance in atomic potential, $t$ is the sample thickness, and $\sigma_e^{(m)}$ is the effective $m$-pass elastic scattering cross section for scatterers with number density $n_s$.
The probability of inelastic scattering is given by $tn_s\sigma_i^{(m)}=1-\eta^m$, so we have
\begin{equation}
    \xi_{\text{MP}}=\frac{4}{\mean{\Delta V^2}}\zeta^{(m)}\ ,
\end{equation}
where 
\begin{equation}
    \zeta^{(m)}=\eta^m\frac{\sigma_e^{(m)}}{\sigma_i^{(m)}}
\end{equation}
is the MP generalization of the information coefficient introduced by Peet et al \cite{Peet2019}. It is a convenient figure of merit because it depends only on material properties rather than the particular configuration of the sample. It also has a straightforward interpretation: when $\zeta^{(m)}>1$, each measurement has a greater chance of detecting a scatterer (probability $\eta^m\sigma_e^{(m)}$) than damaging it (probability $\sigma_i^{(m)}$). If we imagine the scatterer acts like an Elitzer-Vaidman bomb\cite{Elitzur1993} and is destroyed by a single inelastic event, then, assuming post-selection is possible (by registering inelastically scattered electrons), the probability of a counterfactual (guaranteed damage-free) detection is
\begin{equation}
    P_{\text{CF}}=\frac{\zeta^{(m)}}{1+\zeta^{(m)}}\ .
\end{equation}

\begin{figure*}
    \centering
    \includegraphics[width=\textwidth]{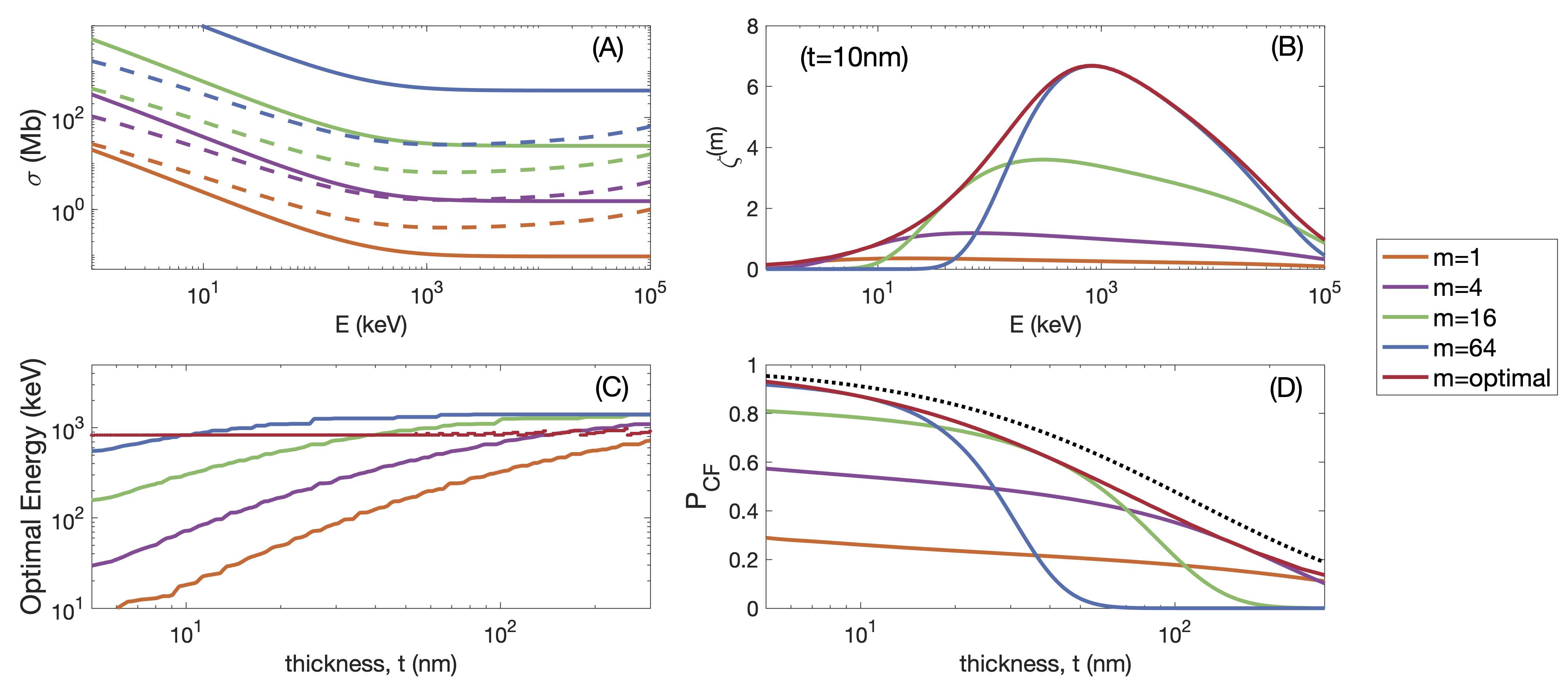}
    \caption{(A) Elastic (solid) and inelastic (dashed) scattering cross sections vs. beam energy for $m$ passes through a thin ($\eta\sim 1$) sample. Notice $\sigma_e^{(m)}\gtrsim\sigma_i^{(m)}$ for $m\gtrsim4$. (B) Information coefficient, $\zeta^{(m)}$, vs. beam energy for various $m$ assuming a sample thickness of $t=\SI{10}{\nm}$. The optimal electron energy is higher for larger $m$. (C) Beam energy which maximizes the information coefficient as a function of sample thickness. Regardless of the sample thickness, the optimal energy (when jointly optimized with $m$) is near \SI{830}{\kilo\electronvolt}. (D) Probability of a successful counterfactual detection of a scatterer at the optimal beam energy for various sample thicknesses. The dotted line represents the quantum limit.}
    \label{fig:InfoCoef}
\end{figure*}

In Fig. \ref{fig:InfoCoef}, we plot various quantities associated with the multi-pass information coefficient. Fig. \ref{fig:InfoCoef} (A) shows the effective elastic cross section $\sigma_e^{(m)}$ and effective inelastic cross section $\sigma_i^{(m)}$ vs. beam energy for amorphous carbon. We calculate $\sigma_e^{(1)}$ using a theoretical relationship proposed by Lamgmore and Smith\cite{Langmore1992}. Values for $\sigma_i^{(1)}$ are taken from the Estar database\cite{Estar}. In (B), we calculate $\zeta^{(m)}$ vs. beam energy for a sample \SI{10}{\nm} thick. In (C) we show the energy which maximizes $\zeta^{(m)}$. The optimal energy $E\sim\SI{830}{\kilo\electronvolt}$ seems to be mostly independent of $t$. Finally, (D) shows the probability of a successful counterfactual detection of a scatterer. To obtain $P_{\text{CF}}>$0.75(0.9) using a MP measurement, the sample thickness must be $t<$\SI{22(7)}{\nm}. Using a quantum-limited measurement, the requirement is $t<$\SI{33(12)}{\nm}.

We can also use a MP measurement to increase the dose efficiency of phase object discrimination. An optimal single pass measurement for discriminating between output wavefunctions $\psi_A$ and $\psi_B$ has success probability equal to the Helstrom Bound\cite{Helstrom}
\begin{equation}
    P_{s}=\frac{1}{2}+\frac{1}{2}\sqrt{1-4P_AP_B\left|\braket{\psi_A}{\psi_B}\right|^2}\ ,
\end{equation}
where $P_A$ and $P_B$ are the \textit{a priori} probabilities of $A$ and $B$.
Taking into account the finite transmissivity of the sample, the MP generalization Helstrom Bound is
\begin{align*}
    P_s^{(m)}&=\sum_{N'=0}^N \eta^{mN'}(1-\eta^m)^{N-N'}{N \choose N'}P_s^{(m)}(N')\\
    P_s^{(m)}(N')&=\frac{1}{2}+\frac{1}{2}\sqrt{1-4P_AP_B\left|\braket{\psi_A^{(m)}{}^{\otimes N'}}{\psi_B^{(m)}{}^{\otimes N'}}\right|^2}\\
    &=\frac{1}{2}+\frac{1}{2}\sqrt{1-4P_AP_B\cos^{2N'}(\beta^{(m)})}\ .
\end{align*}

As an example, suppose we are using a Mach-Zehnder interferometer to categorize an atom as being either species $A$ or $B$ (with prior probabilities $P_A=P_B=0.5$). We'll assume the difference in the phase shifts caused by the two species is $\Delta\phi=0.05$ and also that $\eta=0.98$. Then we can achieve $P_{\text{success}}\gtrsim 0.9$ using either $N=418$ single pass measurements or $N=3$ electrons passed through the sample $m=14$ times each. The single pass method causes, on average, 8.4 inelastic interactions while the MP method causes 0.7 inelastic interactions.

\section{How good can QEM become?}

In his 2016 commentary\cite{howgood} `How good can cryo-EM become?', Glaeser discusses improvements in sample preparation, cameras, and phase contrast techniques that could bring cryo-EM performance to the estimated theoretical limit for a traditional microscope column. He also wondered whether the various proposals for QEMs, which at the time made vague promises about their potential advantage, could save ``true single-particle electron microscopy". Based on our analysis, quantum-limited cryo-EM would yield 10-20 times more information per dose, or 3-4 times better resolution. This would be a significant advancement, but still insufficient for damage free imaging of a single molecule at atomic resolution. However we have made a number of assumptions which could hide the potential for further improvement.

We have not yet emphasized the potentially significant differences between measuring a single phase shift and measuring many simultaneously. In a MP measurement, a transformation could be applied between each pass, modulating the intensity of the beam in a way which could depend on previous interactions with the sample. This `incremental phase contrast' could be especially important in an application like cryo-EM, where the allowable dose can vary throughout the field of view. The structure of the vitreous ice background, for example, need not be preserved. Additionally, the intensity pattern of the initial probe could also be adaptively adjusted between each measurement. This would be similar to adaptively adjusting the dwell time in scanning TEM, which is estimated\cite{adaptiveSTEM} to reduce the required dose by a factor of 10.

Another assumption worth examining is that inelastically-scattered electrons carry no useful information. As discussed in the previous section, post-selection by detection of inelastically-scattered electrons could reveal when and where the sample has been damaged. This would make it possible to expose a sample region of size $R$ precisely until it has been damaged. Taking advantage of this information would require a detailed model of the possible damage processes to determine an appropriate size for $R$.

As an example, suppose it takes 10 counter-factual detections to identify a region. Using the data from Fig \ref{fig:InfoCoef}, a \SI{20}{\nm}-thick sample has a $P_{\text{CF}}=0.72$(0.8) for optimal MP(IFM) and $P_{\text{CF}}=0.23$ for a single pass. Then the chance of a non-destructive MP(IFM) identification is $(P_{\text{CF}})^{10}=4\%(10\%)$ compared to $4\times10^{-5}\%$ for a single pass.

In addition to acting as heralds of damage, inelastically-scattered electrons could provide extra image contrast. While they may be unable to directly contribute to the phase contrast image, it is possible they could be filtered out and measured with a different protocol. For example, they could contribute to a darkfield image, perhaps after being deflected to a second detector. Assuming mean free paths $\lambda_{\text{ice}}\approx\SI{300}{\nm}$ and $\lambda_{\text{protein}}\approx\SI{150}{\nm}$ at beam energy \SI{300}{\keV}\cite{Vulovic2013}, the scattering contrast for a \SI{10}{\nm} thick protein is less than 2\% for 1 pass and 26\% for 16 passes. In addition, when the electron source is subject to Poisson noise (contrary to our assumption in section \ref{sec:QFI}), detection of inelastic electrons could mitigate the extra error\cite{CondReIllum}.

Many of these more speculative techniques would require advances in both hardware and theory. Perhaps in the distant future electron microscopes will use deterministic sources of arbitrary multi-electron quantum states. However, as we have discussed, we can mostly saturate the quantum limit with a classical probe. In order to do that, we would need fast-switching electron optics in order to build a MP TEM column. A number of designs have been proposed for electron resonators using gated electron mirrors \cite{10keVMTEM2019,Turchetti2019,klopfer2021}. Another critical electron optic is a spatial phase modulator, variations of which are currently being demonstrated\cite{PPP,laserPP}. 

Compared to photon microscopy, which is subject to an endless proliferation of design, innovation in electron microscopy is slow. However, the perspective afforded by quantum metrology could motivate more ambitious microscope schemes. There may also be a reciprocal effect on the field of quantum metrology, which could discover a rich new space of applications where damage (rather than probe size) is the limiting resource.

\section{Acknowledgments}
This work was done as part of the Quantum Electron Microscope collaboration funded by the Gordon and Betty Moore foundation. A.J.B. acknowledges support from the Stanford Graduate Fellowship and from the National Science Foundation Graduate Research Fellowship Program under Grant No. 1656518.

\bibliographystyle{unsrt}
\bibliography{main}

\begin{thebibliography}{10}

\bibitem{dubochet1988cryo}
Jacques Dubochet, Marc Adrian, Jiin-Ju Chang, Jean-Claude Homo, Jean Lepault,
  Alasdair~W McDowall, and Patrick Schultz.
\newblock Cryo-electron microscopy of vitrified specimens.
\newblock {\em Quarterly Reviews of Biophysics}, 21(2):129--228, 1988.

\bibitem{frank2006}
Joachim Frank.
\newblock {\em Three-dimensional electron microscopy of macromolecular
  assemblies: visualization of biological molecules in their native state}.
\newblock Oxford university press, 2006.

\bibitem{cryoEMStrucBioreview2015}
Xiao-Chen Bai, Greg McMullan, and Sjors~HW Scheres.
\newblock How cryo-{EM} is revolutionizing structural biology.
\newblock {\em Trends in Biochemical Sciences}, 40(1):49--57, 2015.

\bibitem{SPA}
Richard Henderson.
\newblock The potential and limitations of neutrons, electrons and x-rays for
  atomic resolution microscopy of unstained biological molecules.
\newblock {\em Quarterly Reviews of Biophysics}, 28(2):171–193, 1995.

\bibitem{rickgauer2017}
J.~P. Rickgauer, N.~Grigorieff, and W.~Denk.
\newblock Single-protein detection in crowded molecular environments in
  cryo-{EM} images.
\newblock {\em eLife}, 6:e25648, 2017.

\bibitem{rickgauer2020}
J.~P. Rickgauer, H.~Choi, J.~Lippincott-Schwartz, and W.~Denk.
\newblock Label-free single-instance protein detection in vitrified cells.
\newblock {\em BioRxiv}, 2020.

\bibitem{QuantumMZI_2002}
Hwang Lee, Pieter Kok, and Jonathan~P Dowling.
\newblock A quantum rosetta stone for interferometry.
\newblock {\em Journal of Modern Optics}, 49(14-15):2325--2338, 2002.

\bibitem{QuantMeas_2004}
Vittorio Giovannetti, Seth Lloyd, and Lorenzo Maccone.
\newblock Quantum-enhanced measurements: beating the standard quantum limit.
\newblock {\em Science}, 306(5700):1330--1336, 2004.

\bibitem{QL}
B.~Escher, R.~Filho, and Liema Davidovich.
\newblock General framework for estimating the ultimate precision limit in
  noisy quantum-enhanced metrology.
\newblock {\em Nature Physics}, 7, 2012.

\bibitem{multiphaseEstPRL13}
Peter~C Humphreys, Marco Barbieri, Animesh Datta, and Ian~A Walmsley.
\newblock Quantum enhanced multiple phase estimation.
\newblock {\em Physical Review Letters}, 111(7):070403, 2013.

\bibitem{NOONmicroscopyPRL14}
Yonatan Israel, Shamir Rosen, and Yaron Silberberg.
\newblock Supersensitive polarization microscopy using {NOON} states of light.
\newblock {\em Physical Review Letters}, 112(10):103604, 2014.

\bibitem{QuantumenhancedNM}
Catxere~A. Casacio, L.~Madsen, Alex Terrasson, M.~Waleed, Kai Barnscheidt,
  B.~Hage, Michael Taylor, and W.~Bowen.
\newblock Quantum-enhanced nonlinear microscopy.
\newblock {\em Nature}, 594 7862:201--206, 2021.

\bibitem{FermionNOON_PRA14}
Fabio Benatti, Roberto Floreanini, and Ugo Marzolino.
\newblock Entanglement in fermion systems and quantum metrology.
\newblock {\em Physical Review A}, 89(3):032326, 2014.

\bibitem{Zhao2021}
Zhexin Zhao, Xiao-Qi Sun, and Shanhui Fan.
\newblock Quantum entanglement and modulation enhancement of
  free-electron--bound-electron interaction.
\newblock {\em Physical Review Letters}, 126:233402, 2021.

\bibitem{Hayun2021}
A.~Hayun, O.~Reinhardt, Jonathan Nemirovsky, Aviv Karnieli, N.~Rivera, and
  I.~Kaminer.
\newblock Shaping quantum photonic states using free electrons.
\newblock {\em Science Advances}, 7:eabe4270, 2021.

\bibitem{Okamoto2013}
Hiroshi Okamoto and Yukinori Nagatani.
\newblock Entanglement-assisted electron microscopy based on a flux qubit.
\newblock {\em Applied Physics Letters}, 104, 2013.

\bibitem{ElitzurVaidman1993}
Avshalom~C Elitzur and Lev Vaidman.
\newblock Quantum mechanical interaction-free measurements.
\newblock {\em Foundations of Physics}, 23(7):987--997, 1993.

\bibitem{Vaidman2001}
Lev Vaidman.
\newblock The meaning of the interaction-free measurements.
\newblock {\em Foundations of Physics}, 33, 2001.

\bibitem{kwiat1995}
Paul Kwiat, Harald Weinfurter, Thomas Herzog, Anton Zeilinger, and Mark~A
  Kasevich.
\newblock Interaction-free measurement.
\newblock {\em Physical Review Letters}, 74(24):4763, 1995.

\bibitem{misra1977zeno}
Baidyanath Misra and EC~George Sudarshan.
\newblock The {Zeno’s} paradox in quantum theory.
\newblock {\em Journal of Mathematical Physics}, 18(4):756--763, 1977.

\bibitem{quantumZeno1990}
Wayne~M Itano, Daniel~J Heinzen, John~J Bollinger, and David~J Wineland.
\newblock Quantum {Zeno} effect.
\newblock {\em Physical Review A}, 41(5):2295, 1990.

\bibitem{Putnam2009}
William~P. Putnam and Mehmet~Fatih Yanik.
\newblock Noninvasive electron microscopy with interaction-free quantum
  measurements.
\newblock {\em Physical Review A}, 80:040902, 2009.

\bibitem{Thomas2014}
Sebastian Thomas, C.~Kohstall, P.~Kruit, and P.~Hommelhoff.
\newblock Semitransparency in interaction-free measurements.
\newblock {\em Physical Review A}, 90:053840, 2014.

\bibitem{og_qemdesign}
P.~Kruit, R.~G. Hobbs, C-S. Kim, Y.~Yang, V.~R. Manfrinato, J.~Hammer,
  S.~Thomas, P.~Weber, B.~Klopfer, C.~Kohstall, T.~Juffmann, M.~A. Kasevich,
  P.~Hommelhoff, and K.~K. Berggren.
\newblock Designs for a quantum electron microscope.
\newblock {\em Ultramicroscopy}, 164:31--45, 2016.

\bibitem{E-IFM2021}
Amy~E. Turner, Cameron~W. Johnson, Pieter Kruit, and Benjamin~J. McMorran.
\newblock Interaction-free measurement with electrons.
\newblock {\em Physical Review Letters}, 127:110401, 2021.

\bibitem{MPoptimality2}
Vittorio Giovannetti, Seth Lloyd, and Lorenzo Maccone.
\newblock Quantum metrology.
\newblock {\em Physical Review Letters}, 96:010401, 2006.

\bibitem{MPoptimality1}
V.~Giovannetti, S.~Lloyd, and L.~Maccone.
\newblock Advances in quantum metrology.
\newblock {\em Nature Photonics}, 5:222--229, 2011.

\bibitem{MPoptimality3}
Alfredo Luis.
\newblock Phase-shift amplification for precision measurements without
  nonclassical states.
\newblock {\em Physical Review A}, 65:025802, 2002.

\bibitem{MPinterferometer}
B~Higgins, D~Berry, S~Bartlett, Howard Wiseman, and G~Pryde.
\newblock Entanglement-free heisenberg-limited phase estimation.
\newblock {\em Nature}, 450:393--6, 2007.

\bibitem{MP_NatComm2016}
Thomas Juffmann, Brannon~B Klopfer, Timmo~LI Frankort, Philipp Haslinger, and
  Mark~A Kasevich.
\newblock Multi-pass microscopy.
\newblock {\em Nature Communications}, (1):1--5, 2016.

\bibitem{MPsim_SciRep2017}
Thomas Juffmann, Stewart~A Koppell, Brannon~B Klopfer, Colin Ophus, Robert~M
  Glaeser, and Mark~A Kasevich.
\newblock Multi-pass transmission electron microscopy.
\newblock {\em Scientific Reports}, 7(1):1--7, 2017.

\bibitem{NOONdiscrimination}
A.~Ac\'{\i}n.
\newblock Statistical distinguishability between unitary operations.
\newblock {\em Physical Review Letters}, 87:177901, 2001.

\bibitem{MPdiscrimination}
R.~Duan, Y.~Feng, and M.~Ying.
\newblock Entanglement is not necessary for perfect discrimination between
  unitary operations.
\newblock {\em Physical Review Letters}, 98 10:100503, 2007.

\bibitem{Okamoto2006}
Hiroshi Okamoto, Tatiana Latychevskaia, and Hans-Werner Fink.
\newblock A quantum mechanical scheme to reduce radiation damage in electron
  microscopy.
\newblock {\em Applied Physics Letters}, 88(16):164103, 2006.

\bibitem{qsort}
F.~Troiani, E.~Rotunno, S.~Frabboni, R.~B.~G. Ravelli, P.~J. Peters, E.~Karimi,
  and V.~Grillo.
\newblock Efficient molecule discrimination in electron microscopy through an
  optimized orbital angular momentum sorter.
\newblock {\em Physical Review A}, 102:043510, 2020.

\bibitem{Holevo1982}
Holevo~A S.
\newblock {\em Probabilistic and Statistical Aspects of Quantum Theory}.
\newblock Amsterdam: North- Holland, 1982.

\bibitem{Helstrom1967}
C~W Helstrom.
\newblock {Minimum mean-squared error of estimates in quantum statistics}.
\newblock {\em Physics Letters A}, 25(2):101--102, 1967.

\bibitem{Braunstein1994}
Samuel~L. Braunstein and Carlton~M. Caves.
\newblock Statistical distance and the geometry of quantum states.
\newblock {\em Physical Review Letters}, 72:3439--3443, 1994.

\bibitem{CTF}
L.~Reimer and H.~Kohl.
\newblock {\em Transmission Electron Microscopy}.
\newblock Springer Science, 2008.

\bibitem{schattschneider200Inelcoherence}
Peter Schattschneider and Wolfgang~SM Werner.
\newblock Coherence in electron energy loss spectrometry.
\newblock {\em Journal of electron spectroscopy and related phenomena},
  143(2-3):81--95, 2005.

\bibitem{herring2005EELShologratphy}
RA~Herring.
\newblock Energy-filtered electron-diffracted beam holography.
\newblock {\em Ultramicroscopy}, 104(3-4):261--270, 2005.

\bibitem{egerton2017EELSloc}
RF~Egerton.
\newblock Scattering delocalization and radiation damage in {STEM-EELS}.
\newblock {\em Ultramicroscopy}, 180:115--124, 2017.

\bibitem{lossyQE}
Rafał Demkowicz-Dobrzański, Uwe Dorner, B.~J. Smith, Jeff~S Lundeen, Wojciech
  Wasilewski, Konrad Banaszek, and Ian~A. Walmsley.
\newblock Quantum phase estimation with lossy interferometers.
\newblock {\em Physical Review A}, 80:013825, 2009.

\bibitem{Birchall2017}
Patrick~M. Birchall, Jeremy~L. O'Brien, Jonathan C.~F. Matthews, and Hugo
  Cable.
\newblock Quantum-classical boundary for precision optical phase estimation.
\newblock {\em Physical Review A}, 96:062109, 2017.

\bibitem{FermionSqueezing}
K.~Svozil.
\newblock Squeezed fermion states.
\newblock {\em Physical Review Letters}, 65:3341--3343, 1990.

\bibitem{2nmRes}
Robert Glaeser and Richard Hall.
\newblock Reaching the information limit in cryo-{EM} of biological
  macromolecules: Experimental aspects.
\newblock {\em Biophysical Journal}, 100:2331--7, 2011.

\bibitem{Peet2019}
M.~Peet, R.~Henderson, and C.~J. Russo.
\newblock The energy dependence of contrast and damage in electron
  cryomicroscopy of biological molecules.
\newblock {\em Ultramicroscopy}, 203:125 -- 131, 2019.

\bibitem{Elitzur1993}
A.C. Elitzur and L.~Vaidman.
\newblock Quantum mechanical interaction-free measurements.
\newblock {\em Foundations of Physics}, 23:987–997, 1993.

\bibitem{Langmore1992}
John~P Langmore and Michael~F Smith.
\newblock Quantitative energy-filtered electron microscopy of biological
  molecules in ice.
\newblock {\em Ultramicroscopy}, 46(1-4):349--373, 1992.

\bibitem{Estar}
M.J. Berger, J.S. Coursey, M.A. Zucker, and J.~Chang.
\newblock Stopping-power and range tables for electrons, protons, and helium
  ions, July 2017.

\bibitem{Helstrom}
Carl~W. Helstrom.
\newblock {\em Quantum detection and estimation theory}.
\newblock Academic Press New York, 1976.

\bibitem{howgood}
Robert Glaeser.
\newblock How good can cryo-{EM} become?
\newblock {\em Nature methods}, 13:28--32, 2015.

\bibitem{adaptiveSTEM}
Tim Dahmen, Michael Engstler, Christoph Pauly, Patrick Trampert, Niels Jonge,
  Frank Mücklich, and Philipp Slusallek.
\newblock Feature adaptive sampling for scanning electron microscopy.
\newblock {\em Scientific Reports}, 6:25350, 2016.

\bibitem{Vulovic2013}
Miloš Vulović, Raimond~B.G. Ravelli, Lucas~J. {van Vliet}, Abraham~J. Koster,
  Ivan Lazić, Uwe Lücken, Hans Rullgård, Ozan Öktem, and Bernd Rieger.
\newblock Image formation modeling in cryo-electron microscopy.
\newblock {\em Journal of Structural Biology}, 183(1):19--32, 2013.

\bibitem{CondReIllum}
Akshay Agarwal, Karl~K. Berggren, Yuri~J. van Staaden, and Vivek~K. Goyal.
\newblock Reduced damage in electron microscopy by using interaction-free
  measurement and conditional reillumination.
\newblock {\em Physical Review A}, 99:063809, 2019.

\bibitem{10keVMTEM2019}
Stewart~A Koppell, Marian Mankos, Adam~J Bowman, Yonatan Israel, Thomas
  Juffmann, Brannon~B Klopfer, and Mark~A Kasevich.
\newblock Design for a 10 kev multi-pass transmission electron microscope.
\newblock {\em Ultramicroscopy}, 207:112834, 2019.

\bibitem{Turchetti2019}
Marco Turchetti, Chung-Soo Kim, Richard Hobbs, Yujia Yang, Pieter Kruit, and
  Karl~K. Berggren.
\newblock Design and simulation of a linear electron cavity for quantum
  electron microscopy.
\newblock {\em Ultramicroscopy}, 199:50--61, 2019.

\bibitem{klopfer2021}
Brannon~B Klopfer, Stewart~A Koppell, Adam~J Bowman, Yonatan Israel, and Mark~A
  Kasevich.
\newblock Fast pulse shaping for a novel gated electron mirror.
\newblock {\em Review of Scientific Instruments}, 92(4):043705, 2021.

\bibitem{PPP}
Jo~Verbeeck, Armand Béché, Knut Müller-Caspary, Giulio Guzzinati, Minh
  Luong, and Martien Hertog.
\newblock Demonstration of a 2 × 2 programmable phase plate for electrons.
\newblock {\em Ultramicroscopy}, 190, 2017.

\bibitem{laserPP}
Osip Schwartz, Jeremy Axelrod, Sara Campbell, Carter Turnbaugh, Robert Glaeser,
  and Holger Müller.
\newblock Laser phase plate for transmission electron microscopy.
\newblock {\em Nature Methods}, 16:1016--1020, 2019.

\end{thebibliography}

\end{document}